\documentclass[12pt,preprint]{aastex}

\usepackage{epsfig}
\slugcomment{Accepted to the Astrophysical Journal Letters on 2/12/14}

\shorttitle{The Phases of Water Ice in the Solar Nebula}
\shortauthors{Fred J. Ciesla}

\begin{document}
%\pdfgraphics

%% LaTeX will automatically break titles if they run longer than
%% one line. However, you may use \\ to force a line break if
%% you desire.

\title{The Phases of Water Ice in the Solar Nebula}

\author{F. J. Ciesla\altaffilmark{1}}
\affil{Department of the Geophysical Sciences, The University of Chicago, 5734 South Ellis Avenue, Chicago, IL 60637}
\newpage

\begin{abstract}

Understanding the phases of water ice that were present in the solar nebula has implications for understanding cometary and planetary compositions as well as internal evolution of these bodies.  Here we show that amorphous ice formed more readily than previously recognized, with formation at temperatures $<$70 K being possible under protoplanetary disk conditions.  We further argue that photodesorption and freeze-out of water molecules near the surface layers of the solar nebula would have provided the conditions needed for amorphous ice to form.  This processing would be a natural consequence of ice dynamics, and would allow for the trapping of noble gases and other volatiles in water ice in the outer solar nebula.

\end{abstract}


\newpage

\section{Introduction}

Whether water  was present in the outer solar nebula as amorphous or crystalline ice remains a topic of debate.  If amorphous ice formed in the nebula, then it would have been capable of trapping other gases present and could possibly explain the noble gas contents of comets and Jupiter \citep{barnun85,owen99,yokochi12}.    However, \citet{kouchi94} demonstrated that as water ice is expected to condense at temperatures of 120-180 K in the solar nebula, water molecules would have been able to arrange themselves into a crystalline structure.  Because the D/H ratios of water throughout the solar nebula is significantly less than that found in the interstellar medium, it is thought that much of the ice in the nebula saw high temperatures near the young Sun then was carried outward to cooler regions, condensing at these temperatures along the way \citep{mousis00,yang13}.  For example, bringing the D/H ratio of water from the interstellar ratio of $\sim$0.001-0.01 to the cometary values of 1.5-3$\times$10$^{-4}$ \citep{hartogh11} requires that that 70-97\% of water equilibrated with molecular hydrogen (solar D/H $\sim$10$^{-5}$) at high temperatures ($>$500 K) in the solar nebula, before cooling and condensing as ice.  This has led many to argue that amorphous ice formation was precluded in the solar nebula, and that all water ice was crystalline requiring another means of trapping volatiles such as clathrates \citep{kellerjorda01,hersant04}.

The solar nebula is expected to have been a dynamic object, in which gas and dust were pushed around as mass and angular momentum were transported throughout the disk \citep{armitage11}.  As a consequence of this evolution, gas and dust would be transported throughout the disk and exposed to a variety of physical and chemical environments.  Upon passing through such environments, the dust and gas could be altered by chemical reactions, irradiation, or phase transitions.  This was demonstrated in \citet{cieslasandford12}, as icy grains in the outer solar nebula were found to be lofted to high altitudes above the disk midplane and exposed to intense radiation fluxes which would break molecular bonds in the ices and allow the resulting ions and radicals to react to form more complex species.  Thus just because a compound or phase is formed in a protoplanetary disk, it does not mean that original phase is preserved throughout the lifetime of the solar nebula.

In this Letter we revisit the conditions under which the various phases of water ice would form in the solar nebula.  In particular, the conditions for amorphous ice formation in the solar nebula are derived assuming that the ice forms on a substrate other than hexagonally crystalline ice.  We then show that the conditions for amorphous ice formation would have been met as icy grains were irradiated by UV photons at the surface layers of the solar nebula and the resulting photodesorbed water froze-out again at greater depths inside the protoplanetary disk.

\section{Formation Conditions of Ice Phases}

  \citet{kouchi94}  determined that the phase of water ice that forms during condensation is controlled by the ability of condensing water molecules to rearrange themselves on the surface of the substrate.  That is, if water molecules condense at some flux below a critical value, $F<F_{c}$,  the water molecules are able to diffuse across the surface and achieve a minimum energy state, thus forming crystalline ice.  At higher fluxes, $F>F_{c}$, the delivery of water molecules to the surface is so rapid that they cannot rearrange themselves into an ordered state before being locked into place.  This results in the formation of amorphous ice.  A key factor in setting $F_{c}$ is the surface diffusivity of the substrate the water ice lands on: for high surface diffusivity, water molecules can readily rearrange themselves and form crystalline ice, while lower surface diffusivities lead to the formation of amorphous ice.  As surface diffusivity is temperature dependent, amorphous ice is most easily formed at low temperatures.

 Using the surface diffusion for hexagonally crystalline water ice ($I_{h}$), \citet{kouchi94} found that throughout the solar nebula, all water would have frozen out at fluxes that were orders of magnitude below $F_{c}$ (=$D$/$a^{4}$ where $D$ is the surface diffusivity of the substrate and $a$=4.5$\times$10$^{-8}$ cm is the lattice constant of crystalline ice), and thus concluded that all water ice in the solar nebula would be crystalline. 
$I_{h}$ has a relatively high surface diffusivity ($D_{h}$=0.045 exp(-1230/$T$) cm$^{2}$ s$^{-1}$) for ice.  However, in that same study, the authors experimentally determined the surface diffusivity for polycrystalline or cubic crystalline ice (represented by $I_{p}$ here, but often seen as $I_{c}$--we avoid using the subscript $c$ to avoid confusion when talking about $F_{c}$) a phase whose discontinuities in its crystal lattice and uneven surface leads to low surface diffusion rates ($D_{p}$=1.74$\times$10$^{5}$ exp(-4590/$T$) cm$^{2}$ s$^{-1}$) \citep{kouchi94}. $I_{p}$ will form at temperatures $< \sim$200 K as a metastable phase, whereas $I_{h}$ will be the crystalline form at higher temperatures \citep{baragiola03,hobbs}.  As water ice would have formed in the solar nebula at temperatures between 120-180 K (depending on ambient pressure) water ice initially would have been in the form of $I_{p}$ rather than $I_{h}$.

Figure 1 compares the values of $F_{c}$ for the case of water condensing onto $I_{h}$ as done by \citet{kouchi94} and onto $I_{p}$ using the parameters for polycrystalline ice derived from the experiments in that same paper.  These fluxes are plotted as functions of temperature.  For reference, the condensing flux of water vapor in a minimum mass solar nebula \citep[MMSN][]{weidenschilling77,hayashi81} and a disk with a surface density of $\Sigma \left( r \right)$=2000($r$/1 AU)$^{-1}$ and $T \left( r \right)$=150($r$/1 AU)$^{-\frac{3}{7}}$ which represents a slightly more massive, irradiated disk \citep{chiang97}, are also included.  For both disks, the condensing flux is calculated at the disk midplane for the corresponding temperatures, where the flux would be greatest if all water began as vapor.

As can be seen, $F_{c}$ is orders of magnitude higher for $I_{h}$ than $I_{p}$ at all temperatures relevant to the solar nebula.  Whereas $F_{c}$ would have reached low enough values to allow amorphous ice to form at temperatures $<$20 K in the solar nebula  based on the \citet{kouchi94} calculations, amorphous ice may be able to form at temperatures of 70-90 K if one accounts for the fact that $I_{p}$ is likely the original phase for water ice in the disk.

\section{Ice Formation Conditions in the Solar Nebula}

While amorphous ice may be able to form at higher temperatures than originally calculated by \citet{kouchi94}, these temperatures are still significantly below the temperatures where water ice is expected to condense under solar nebula temperatures.  That is, as a gas of solar composition cools, or as a parcel of material migrates outward from the inner regions of the solar nebula, water ice forms at temperatures in the 120-180 K range, leaving negligible water in the vapor phase to condense at lower temperatures where amorphous ice would be expected.

However, a reservoir of cold water vapor has recently been detected in the protoplanetary disk around TW Hydrae by the Herschel Space Observatory \citep{hogerheijde11}.  This water vapor is inferred to be present at the disk surface, with most of the observed emission lines originating 100-200 AU from the central star, where temperatures are $<$50 K.  As thermal desorption would be minimal here, the water vapor is believed to result from the UV photodesorption of water ice covered grains at the disk surface, where 1 out of every $\sim$1000 incident photons liberates a water molecule  \citep[net yield $Y \sim$0.001,][]{oberg09}.  Water molecules liberated in this way, however, would not remain in the gas phase, and would freeze out again at the low temperatures found in the disk in this region (if they are not photo-dissociated themselves, as we discuss further below).

To explore how much water would be liberated and freeze-out again in this manner, we consider the distribution of water vapor in a column of a protoplanetary disk at $r$=30 AU, with a surface density of 60 g cm$^{-2}$ and a temperature of 20 K (values close to the irradiated disk considered above).  Such values would be expected for a typical T-Tauri disk   We take the column to be isothermal; while disk surface layers may reach higher temperatures due to direct irradiation from the central star, the temperatures expected there would be 30-50 K, still below the temperatures where thermal desorption is important and thus these temperature variations are not important for our purposes.  We consider dust grains to be uniformly distributed throughout the column with an average radius of $a$=1 $\mu$m.  Assuming a grain density of $\rho_{g}$=2 g cm$^{-3}$ and dust-to-gas mass ratio of 0.01, the opacity of the column is $\kappa$=37.5 cm$^{2}$ g$^{-1}$.

The  freeze-out flux of water vapor would be given by $F_{FO}$=$n_{H_{2}O} v_{th}$ where $n_{H_{2}O}$ is the ambient density of water molecules and $v_{th}$ the thermal velocity of the water molecules. The photodesorption flux would be given by $F_{PD}$=$G_{0}$10$^{8}Ye^{-\tau}$, where 10$^{8}$ is the flux of UV photons in the interstellar radiation field \citep[in cm$^{-2}$ s$^{-1}$][]{habing68}, $G_{0}$ is a factor which measures how much stronger the incident UV radiation field is than the interstellar field, and $\tau$ is the optical depth into the disk of the region of interest.  Here we assume pure water ice on the grains and ignore other ices for now.  The net flux $F_{net}$=$F_{PD}-F_{FO}$ will be positive, meaning water molecules will be lost from the grains and added to the vapor in the upper regions of the disk where UV fluxes are high (optical depths are low).  Freeze-out and ice formation, will occur when $F_{net} < 0$ deeper in the disk.  Thus the flux at which water is delivered to the surface of a grain in this situation will depend on the time it takes for grains and vapor to migrate from the UV irradiated layer to the shielded disk interior.

 To estimate the typical time for such migration and the corresponding fluxes of water from and onto the grains, the particle tracking methods of \citet{ciesla10mc} were used to track how parcels of materials would migrate through the vertical column of the disk considered here.  The path of a typical parcel is shown in Figure 2  over a 10$^{5}$ year time period.  For the sake of our calculations, we consider the case of the disk with an incident flux of $G_{0}$=10$^{4}$ at 30 AU, which is a typical value for a T Tauri star due to the accretion of mass onto the young star \citep{bergin04}. The radiation was assumed to propagate perpendicular to the disk for simplicity--in reality, the radiation from the central star will be incident at some angle, limiting the depth of penetration to less than found here.  We simulated this effect by increasing the opacity of the column by factors of 5-10, such that the UV dominated region is limited to the very upper surface layers of the disk and found this to have no major effect on our conclusions.   A turbulent parameter of $\alpha$=10$^{-2}$ was assumed.

 The net flux of water to/from the grain surface along the path was found by calculating the net flux as described above at each point along the trajectory of the materials considered, with solid particles and water vapor co-moving.  All particles began at the disk midplane with all water condensed on their surfaces (H$_{2}$O/H$_{2}$ abundance ratio of 2$\times$10$^{-4}$ was used here--uncertainties in this ratio are  on the order of a few, and thus have a small impact on our results).  This net flux is plotted as a function of the amount of vapor residing in the gas in Figure 2.  The positive fluxes correspond to times when photodesorption dominates, leading to water molecules being incorporated into the gas.  As the grain and vapor migrate deeper into the disk, the flux is negative, meaning ice formation occurs.  The magnitude of fluxes at which this ice forms range from $\sim$10$^{5}$-10$^{9}$ molecules cm$^{-2}$ s$^{-1}$ when more than 10\% of the water has been photodesorbed. These fluxes exceed  $F_{c}$ in Figure 1 by orders of magnitude for both the case of ice formation on $I_{h}$ and $I_{p}$ at the 20 K temperature considered here.   The same calculation was repeated for multiple particles and at temperatures ranging from 10-50 K.  In all cases grains were found to reach altitudes where they would lose water due to photodesorption and then see water ice reform at fluxes that exceeded $F_{c}(I_{p})$.  As such fluxes were achieved at higher temperatures, this indicates that the warmer temperatures at higher altitudes would still allow amorphous ice to form.   Thus the water vapor would freeze-out as it moved to lower heights in the disk as amorphous ice, allowing water which began as crystalline ice would be transformed to amorphous ice by the cycling described here.

\section{Discussion}

The results presented here show that even if water ice condensed in the solar nebula at temperatures in the 120-180 K range, where crystalline ice is expected to form, the dynamical evolution of the icy grains in the outer solar nebula would lead this ice to be lost and reformed as amorphous ice.  Formation of amorphous ice in this manner may allow for the trapping of other gaseous species, such as N- and C- bearing species as well as noble gases.  The trapping of such species in ices would allow volatile-rich planetesimals to form in the outer nebula, and may explain the enhanced abundances of these elements relative to hydrogen in the Jovian atmosphere compared to a gas of solar composition \citep{owen99,atreya03}.

An issue that requires future attention is the detailed fate of the liberated water at the upper layers of the protoplanetary disk.  If water molecules are desorbed sufficiently high in the disk, they may be photodissociated by the same UV photons that liberate them from the grains they are on.  The extent to which materials are photodissociated will depend on the amount of time materials spend in the extreme upper layers of the disk.  As shown here, the amount of water surrounding a grain will constantly be changing, with molecules freezing out and then being liberated over and over again as the conditions they exposed to are constantly changing.  Thus there will be a time component to determining the precise fraction of water molecules that are liberated around particular grains that gets photodissociated compared to the bulk water present in a given region.  

However, estimates of the amount of water vapor present in irradiated regions of the disk under steady-state conditions with no dynamic transport give $n_{H_{2}O}$/$n_{H} \sim$2$\times$10$^{-7}$, or 1/1000 the maximum value reached here \citep{hollenbach09}.  Taking this as an upper limit for the amount of water vapor present in the regions of the disk considered would imply freeze-out fluxes that are ~1000$\times$ smaller than calculated here, or 10$^{2}$-10$^{6}$ cm$^{-2}$ s$^{-1}$.  These values still exceed the $F_{c}(I_{p})$  (Fig. 1) for temperatures $<$60 K by orders of magnitude, meaning amorphous ice would still form. Further, even with photodissociation, the resulting species will freeze-out again and reform H$_{2}$O on grain surfaces at very rapid rates \citep{ioppolo08,dulieu10}.  Given the low surface diffusivities of the molecules on $I_{p}$ at the temperatures considered here, when such molecules form they would not be able to reorder themselves, resulting in amorphous ice.
 
 The formation of amorphous ice alone would aid in the continued formation of this phase of water ice: the surface diffusivity of amorphous ice is orders of magnitude smaller than $I_{p}$ \citep{kouchi94}, meaning any water molecules formed or frozen out on it would remain locked in place and unable to rearrange into a crystalline structure.  This would be true whether the amorphous ice formed via direct freeze-out, surface processes, or through irradiation damage.  \citet{letobaratta03} found that crystalline water ice would be amorphized after receiving a dose of UV photons of a few eV per molecule, or a few UV photons per molecule.  The dosages calculated here and found by \citet{cieslasandford12} are beyond those needed for amorphization.  Thus any water ice that forms as a result of photodesorption in the outer solar nebula would likely form on a very rough substrate with limited surface diffusivity, leading the water to freeze out in the amorphous phase.

Should we thus expect all water ice in the solar nebula and protoplanetary disks to become amorphous?  Not necessarily.  Just as we have shown that cycling of this type could lead crystalline ice to be transformed to amorphous ice, amorphous ice could be transformed to crystalline as it migrates through the protoplanetary disk.  At temperatures near $\sim$100 K, amorphous ice undergoes a spontaneous phase transition, where molecules rearrange themselves to form crystalline ice.  Thus as ice particles migrate radially in a protoplanetary disk and reach higher temperatures, the amorphous ice may be lost.  Trapped volatiles which resided on the grain surface when ice molecules froze out on top of them may be retained, despite the phase transition \citep{viti04, collings04}, meaning that crystalline ices with trapped volatiles may also be present in such disks.  Determining the relative abundances of each phase of ice requires a detailed investigation of the radial transport of icy grains throughout the evolution of the solar nebula, which should be the goal of future work.

%\emph{Acknowledgments} The author is grateful for stimulating and insightful conversations with Ted Bergin and Reika Yokochi. This work was supported by NASA Origins Grant NNX08AY47G awarded to FJC.

\bibliographystyle{apj}

\newpage
\begin{figure}
\includegraphics[angle=90,width=6.in]{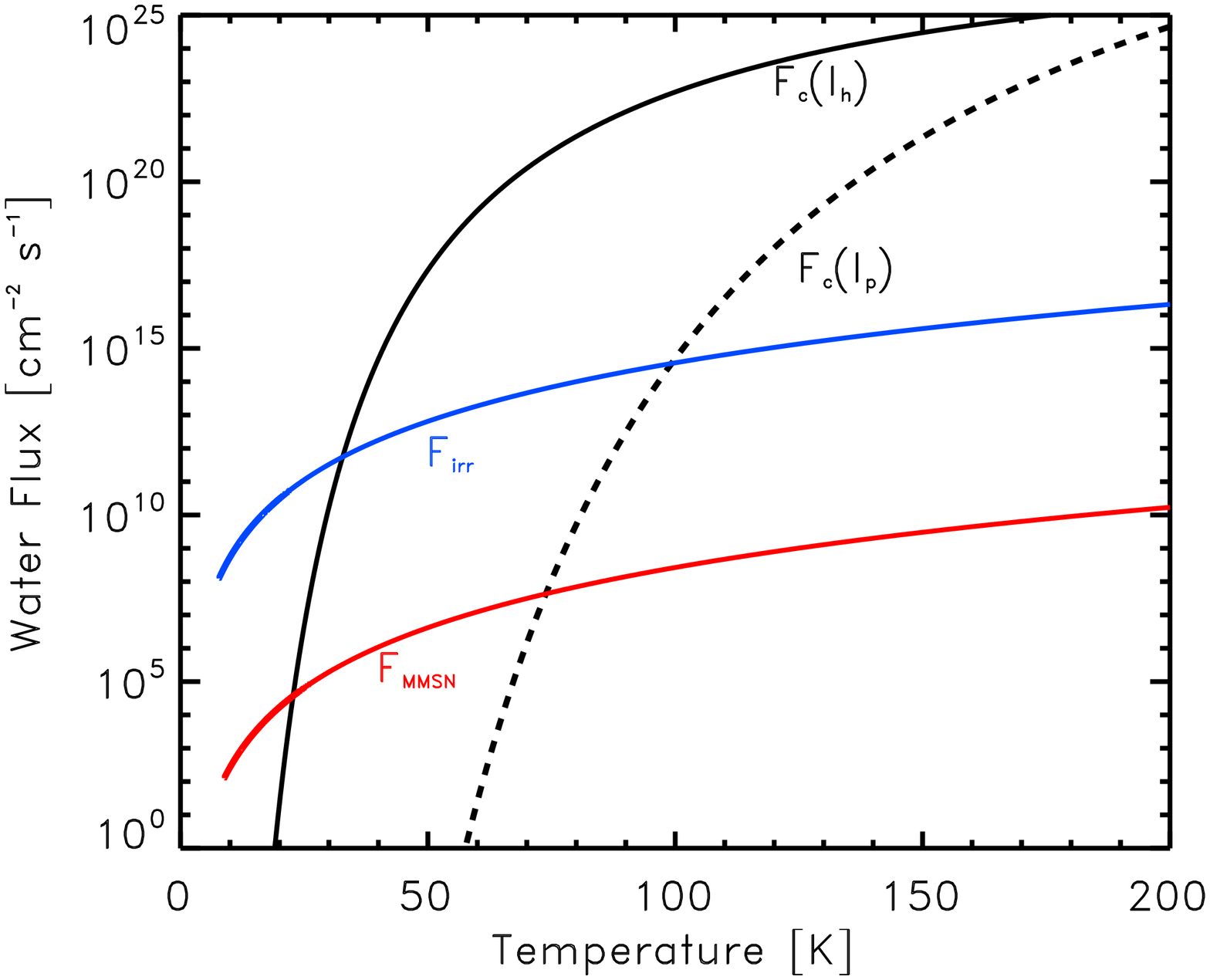}
\caption{Black lines show the critical flux ($F_{c}$) for amorphous ice formation versus temperature for a substrate of hexagonal ice (I$_{h}$, solid) and polycrystalline ice ($I_{p}$, dashed) from \citep{kouchi94}.  The flux of condensing water at the disk midplane for a Minimum Mass Solar Nebula (MMSN, red) and for an irradiated disk with surface density $\Sigma$=2000($r$/1 AU)$^{-1}$ g cm$^{-2}$ (irr, blue) are also shown.}
\end{figure}

\newpage
\begin{figure}
\includegraphics[angle=90,width=3.in]{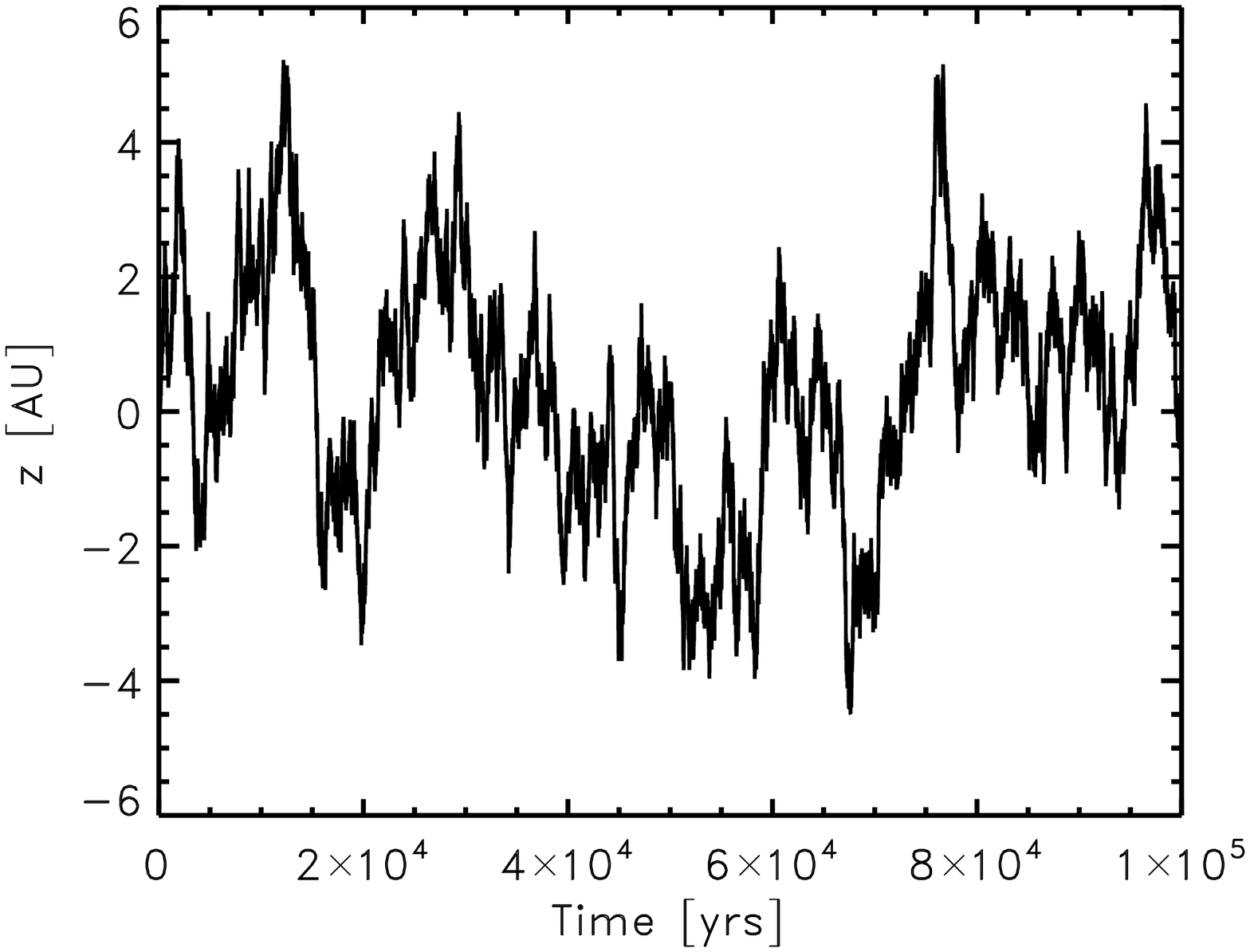}
\includegraphics[angle=90,width=3.in]{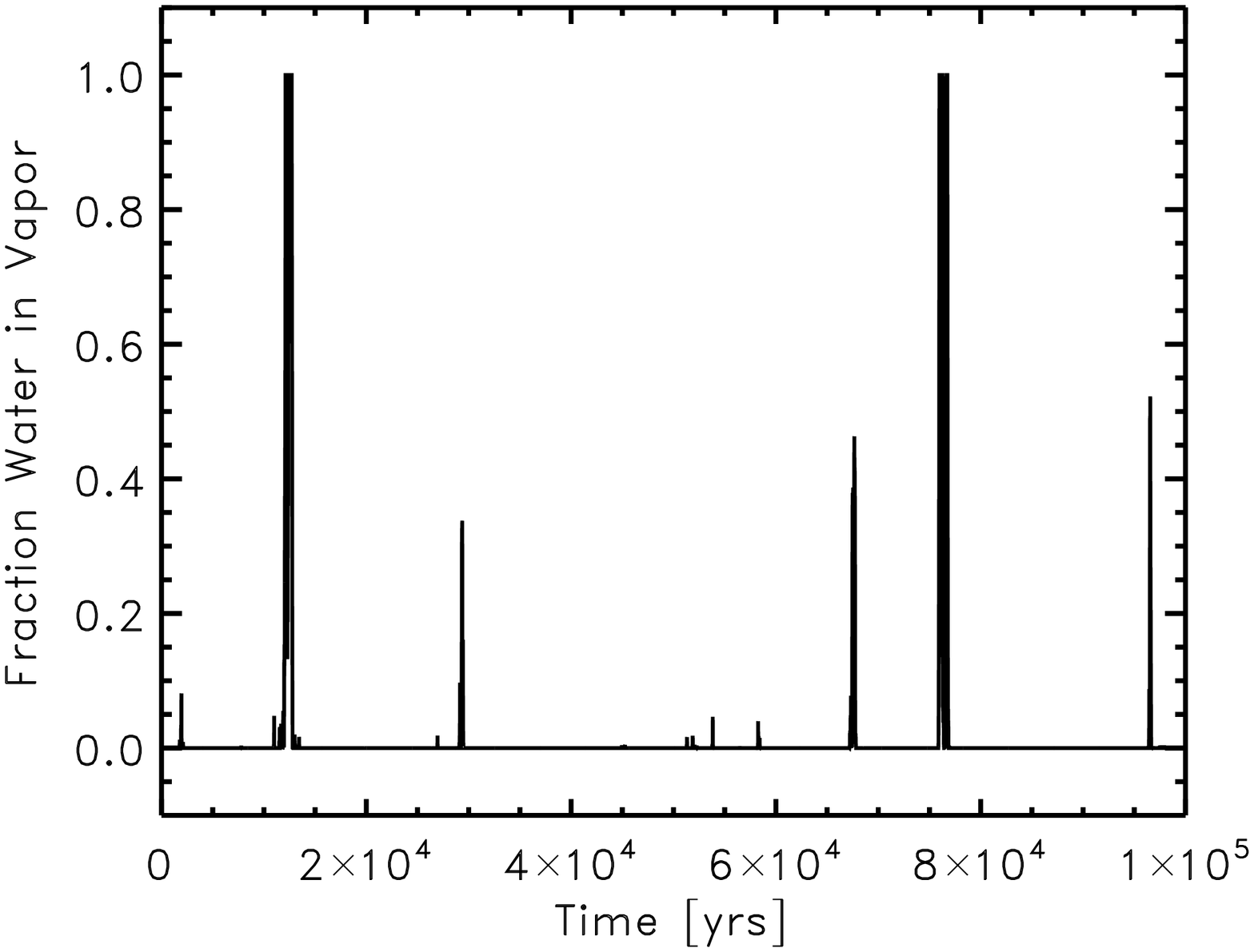}
\caption{\emph{Left:} Typical vertical path of a parcel of solar nebula materials at $r$=30 AU. \emph{Right:} The fraction of water residing in the vapor phase for the particle shown on the left.  Note the greatest fraction of water in the vapor corresponds to times when the particle is lofted to very high altitudes.  Residence in the photodesorption region range from less than 0.1 to a few years.}
\end{figure}

\newpage
\begin{figure}
\includegraphics[angle=90,width=6.in]{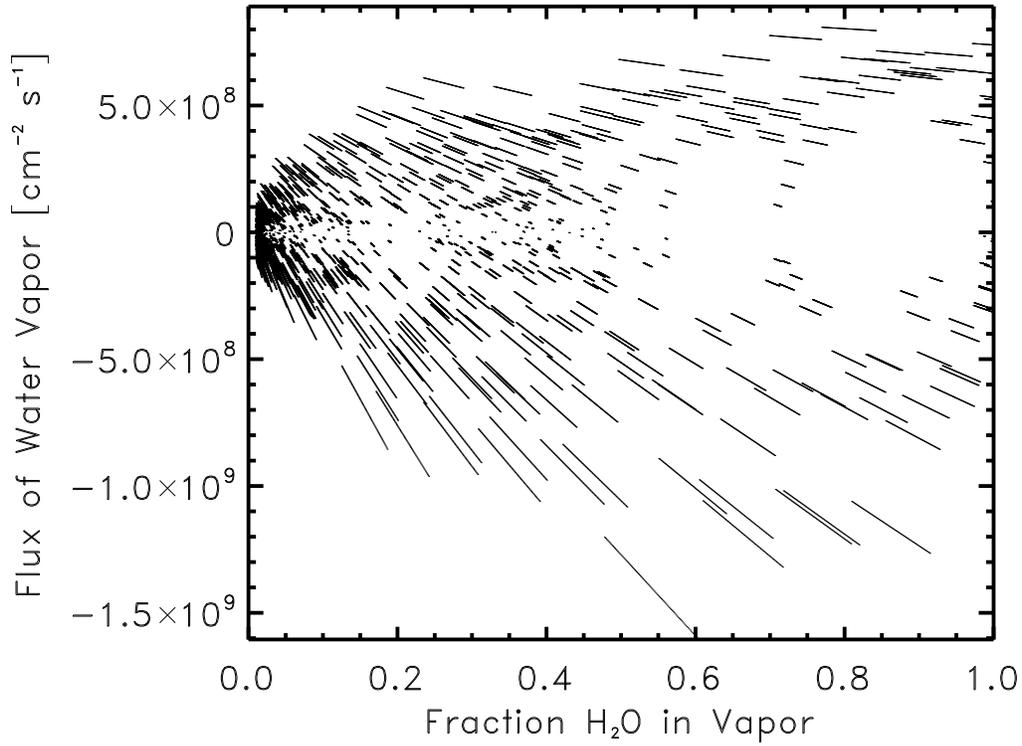}
\caption{Flux of water vapor for the parcel of material whose path is shown in Figure 2 as a function of the amount of water found in the vapor.  Negative fluxes indicate freeze-out of water.  Freeze-out fluxes are in the range $\sim$10$^{5}$-10$^{9}$ cm$^{-2}$ s$^{-1}$.  These fluxes may be 1000$\times$ smaller where photodissociation of water is important.  Still, these fluxes exceed $F_{c} \left(I_{p}\right)$ as shown in Figure 1, indicating amorphous ice would then form.}
\end{figure}

\end{document}